\documentclass[10pt]{iopart}
\usepackage{graphicx}

\usepackage{cite}
\usepackage[linkcolor=blue,citecolor=blue,colorlinks=true,urlcolor=blue]{hyperref}
\expandafter\let\csname equation*\endcsname\relax

\expandafter\let\csname endequation*\endcsname\relax

\usepackage{amssymb,amsmath}
\usepackage[T1]{fontenc}
\usepackage[utf8]{inputenc}
\newcommand{\eq}[1]{Eq.~(\ref{#1})}
\newcommand{\fig}[1]{Fig.~\ref{#1}}

\begin{document}

\title[Three-electron escape in strongly driven Ne at high intensities]{Three-electron escape in strongly driven Ne at high intensities}
\author{S. J. Praill$^{1}$,
G. P. Katsoulis$^{1}$ and 
A. Emmanouilidou$^{1}$\footnote{Corresponding Author: a.emmanouilidou@ucl.ac.uk}}
\address{$^{1}$Department of Physics and Astronomy, University College London, Gower Street, London WC1E 6BT, United Kingdom}



\vspace{10pt}

\begin{abstract}
We extend  a recently developed three-dimensional semiclassical model to study double and triple ionization of Ne driven by infrared laser pulses at various intensities. This model fully accounts for the Coulomb singularity of each electron with the core, as well as for the interaction of a recolliding electron with a bound electron. The model avoids artificial autoionization by employing effective Coulomb potentials to describe the interaction between bound electrons. Using the extended effective-Coulomb-potential for bound-bound electrons (ECBB) model, we compute 
 triple and double ionization spectra. For instance, we compute the distributions of the sum of the final momenta along the laser field of the escaping electrons. Taking focal volume averaging into account, we find very good agreement  with experimental results, particularly for triple ionization. 
Also, we identify the main pathways of triple and double ionization and explain how these pathways give rise to the main features of the triple and double ionization spectra.
\end{abstract}

%
%
\submitto{\JPB}
%
%
\ioptwocol
%

\section{Introduction}

The interaction of atoms and molecules with intense infrared laser fields gives rise to a plethora of interesting phenomena, such as high-harmonic generation \cite{lewenstein_theory_1994,ferray_multiple-harmonic_1988}, above-threshold ionization \cite{agostini_free-free_1979}, and nonsequential multiple ionization (NSMI) \cite{weber_sequential_2000,talebpour_non-sequential_1997,Becker_2005}. The latter is of fundamental interest since it is mediated by electronic correlation \cite{Dorner2000}, offering valuable insights into the intricate dynamics of ultrafast processes. There are several experimental studies  of NSMI, e.g., in strongly-driven Ar and Ne \cite{PhysRevLett.84.447,Rudenko_2008,PhysRevA.86.043402,Herrwerth_2008,Zrost_2006,PhysRevLett.93.253001,SHIMADA2005221}. However, theoretical studies still constitute a computational challenge especially when considering more than two electrons. Indeed, three-dimensional (3D) \textit{ab initio} quantum-mechanical approaches are generally restricted to treating double ionization in two-electron atoms \cite{Parker_2000,PhysRevLett.96.133001}. In addition, a range of quantum-mechanical \cite{PhysRevA.93.023406,PhysRevA.103.053123} and semiclassical methods \cite{PhysRevA.63.043416,PhysRevA.65.021406,PhysRevA.78.023411} that explicitly incorporate the Coulomb singularity have been developed to investigate double ionization. In contrast, for triple ionization, few theoretical studies exist, with a number of approximations. These investigations of triple ionization encompass classical models formulated with reduced dimensionality \cite{PhysRevA.64.053401} and soft-core Coulomb potentials \cite{PhysRevA.78.013401,PhysRevLett.97.083001,Zhou:10,Tang:13,Ho:07}. They also include quantum-mechanical reduced-dimensionality models \cite{PhysRevA.98.031401,Prauzner_Bechcicki_2021, efimov:21} as well as semiclassical methods that make use of Heisenberg potentials \cite{PhysRevA.104.023113}.

One of the central challenges encountered in quantum-mechanical studies of triple ionization in strongly driven systems is the immense computational resources required. This is one reason for the aforementioned use of quantum mechanical models with reduced dimensionality. Classical and semiclassical techniques require less computational resources. However, they face a different issue,  artificial autoionization. Specifically, a bound electron may approach the core closely enough to experience a strong Coulomb attraction, causing its energy to reach a very negative value. Indeed, in classical mechanics there is no lower energy bound limit, contrary to quantum mechanics.  This electron can then give energy, via Coulomb interaction, to another  bound electron, leading to the later electron's escape.  One way to tackle  artificial autoionization is to soften the Coulomb potential, see for example Refs. \cite{PhysRevLett.97.083001,Zhou:10,Tang:13}. Another approach is including Heisenberg potentials (effective softening) in the Hamiltonian, which mimic the Heisenberg uncertainty principle \cite{PhysRevA.21.834,PhysRevA.104.023113,PhysRevA.105.053119}. Such potentials prevent electrons from having a close encounter with the core and, hence, the acquisition of very negative energy. The problem of this approach lies in the fact that the use of softened Coulomb potentials does not adequately describe the electron–core scattering \cite{Pandit2018,Pandit2017,PhysRevA.107.L041101}; this limitation becomes especially pronounced for high-energy recolliding electrons \cite{Pandit2018,Pandit2017,PhysRevA.107.L041101}.

 We have recently  tackled  artificial autoionization by taking a different approach.  We developed a 3D semiclassical model that includes the Coulomb singularity for atoms \cite{Agapi3electron,PhysRevA.107.L041101} and molecules \cite{PhysRevA.109.033106}. The main premise of our model is that during a recollision the interaction between  the recolliding electron, which is a quasifree electron, and the core as well as the interaction between a quasifree  and a bound electron are the most important interactions. Hence, we treat these latter interactions with full Coulomb potentials. Here, quasifree denotes a recolliding electron or one that escapes or has already escaped into the continuum. However, we use effective Coulomb potentials to describe the interaction of a bound-bound electron pair (ECBB model). That is, we approximate the energy transfer between bound electrons. Consequently, the ECBB model is expected to provide greater accuracy for laser pulse parameters in which multi-electron ionization, resulting from energy transfer between electrons bound in excited states after a recollision, plays a minimal role. If this energy transfer plays more of a role, we expect the ECBB model to be less accurate.
 
 Another sophisticated aspect of the 3D ECBB model is that it classifies an electron as quasifree or bound on the fly during time propagation. To do so,  we use a set of simple criteria detailed in Ref. \cite{Agapi3electron}.  Also, the ECBB model is developed in the nondipole framework allowing for the identification of effects due to the magnetic field of the laser pulse \cite{PhysRevA.108.043111}.  The ECBB semiclassical model has previously yielded very good agreement with experimental results for triple ionization of strongly driven Ne at lower intensities, 1 PW/cm$^2$, 1.3 PW/cm$^2$ and 1.6 PW/cm$^2$ \cite{PhysRevA.107.L041101}. Moreover, the ECBB model, in agreement with quantum mechanical studies, accurately predicts the degree of correlation of three-electron escape dynamics, demonstrated by Dalitz plots \cite{43wt-x129}.

It is unsurprising that semiclassical models incorporating the Coulomb singularity are able to accurately reproduce multiple ionization spectra. We have previously shown that a 3D classical model \cite{PhysRevA.78.023411}, the predecessor of the ECBB model, accurately describes double ionization spectra in strongly driven atoms. These spectra agree  with both experimental and quantum mechanical studies \cite{AhaiScientificReport,PhysRevA.100.043410,Emmanouilidou_2011}. Also,  with another 3D classical model that fully accounts for the Coulomb singularities, we have previously described three- and four-electron escape by single photon absorption \cite{Emmanouilidou_2006,PhysRevA.76.054701}. For photon energies close to the ionization threshold, we showed that, depending on the initial state, the three electrons escape either in a T-shape structure or on an equilateral triangle \cite{PhysRevLett.100.063002}. Only the latter  pattern is predicted by Wannier's law \cite{PhysRev.90.817,Klar_1976,Dimitrijevic_1981}. The T-shape structure, which we have found to be   mediated by a sequence of attosecond collisions \cite{PhysRevA.75.022712},  has been verified quantum mechanically \cite{PhysRevLett.110.063001}.

  Here, we extend the ECBB model to higher intensities. Compared to our previous studies \cite{Agapi3electron,PhysRevA.107.L041101}, this extension involves accounting  for tunneling during propagation in strongly driven atoms. It also involves  modifying the initial conditions for the electron that tunnels at the start of the propagation. In addition, we perform an average over the focal volume \cite{wang_disentangling_2005} to obtain spectra that we can more accurately compare with experiment \cite{Rudenko_2008,Zrost_2006,PhysRevLett.93.253001}. Also, we  add another criterion for labelling an electron as quasifree or bound, necessary at higher intensities of the laser field. These additions to the ECBB model are detailed in Section \ref{Section::Method}.
  
   With this extended ECBB model, we investigate nonsequential triple ionization (NSTI)  and nonsequential double ionization (NSDI) of Ne, driven by an 800 nm, 25 fs laser pulse for a range of intensities. At high intensities of 2 PW/cm$^2$ and 3 PW/cm$^2$, we obtain  NSTI and NSDI spectra accounting for focal volume averaging, such as the probability distribution of the sum of the electron momenta along the laser field. We compare these spectra with experimental results \cite{Rudenko_2008,Zrost_2006} and  find a better agreement for NSTI rather than NSDI.  We also find the main pathways of NSTI and NSDI for a range of intensities and  identify new pathways of multi-electron escape  at higher intensities.



\section{Method} \label{Section::Method}
\subsection{ECBB model}
The Hamiltonian of the three-electron atom is given by
\begin{align}\label{Hamiltonian_effective}
&H = \sum_{i=1}^{4}\frac{\left[\mathbf{\tilde{p}}_{i}- Q_i\mathbf{A}(y_i,t) \right]^2}{2m_i}+\sum_{i=1}^{3}\frac{Q_iQ_4}{|\mathbf{r}_i-\mathbf{r}_4|} \nonumber \\
&+\sum_{i=1}^{2}\sum_{j=i+1}^{3} \left[ 1-c_{i,j}(t)\right]\frac{Q_iQ_j}{|\mathbf{r}_i-\mathbf{r}_j|} +\sum_{i=1}^{2}\sum_{j=i+1}^{3}c_{i,j}(t) \nonumber \\
&\times\Big[V_{\mathrm{eff}}(\zeta_j(t),|\mathbf{r}_{4}-\mathbf{r}_{i}|) + V_{\mathrm{eff}}(\zeta_i(t),|\mathbf{r}_{4}-\mathbf{r}_{j}|)\Big],
\end{align}
where $Q_i$ is the charge, $m_i$ is the mass, $\mathbf{r}_{i}$ is the position vector and $\mathbf{\tilde{p}}_{i}$ is the canonical momentum vector of particle $i$. The mechanical momentum $\mathbf{p}_{i}$ is given by
\begin{equation}
\mathbf{p}_{i}=\mathbf{\tilde{p}}_{i}-Q_i\mathbf{A}(y_i,t),
\end{equation}
where $\mathbf{A}(y,t)$ is the vector potential and $\mathbf{E}(y,t) = - \dfrac{\partial\mathbf{A}(y,t)}{\partial t}$ is the electric field of the laser pulse. We note that, in the ECBB model, we account for the motion of the three electrons and the nucleus at the same time. To derive the effective Coulomb potential that an electron $i$ experiences at a distance $|\mathbf{r}_{4}-\mathbf{r}_{i}|$ from the core (particle 4 with $Q_{4}=3$ a.u.), due to the charge distribution of electron $j$, we approximate the wavefunction of a bound electron $j$ with a 1s hydrogenic wavefunction
\begin{equation}
\psi(\zeta_j,|\mathbf{r}_{4}-\mathbf{r}_{j}|) = \left( \frac{\zeta_j^3}{\pi} \right)^{1/2} e^{-\zeta_j |\mathbf{r}_{4}-\mathbf{r}_{j}|},
\end{equation} 
with $\zeta_j$ the effective charge of particle $j$ \cite{Agapi3electron,PhysRevA.40.6223}. Then, using Gauss's law \cite{Agapi3electron,PhysRevA.40.6223}, one finds that the potential produced due to the charge distribution $-|\psi(\zeta_j,|\mathbf{r}_{4}-\mathbf{r}_{j}|) |^2$ is given by
\begin{equation} \label{eqn::effective_potential}
V_{\mathrm{eff}}(\zeta_j,|\mathbf{r}_{4}-\mathbf{r}_{i}|) =  \dfrac{1-(1+\zeta_j| \mathbf{r}_{4}-\mathbf{r}_{i}|)e^{-2\zeta_j| \mathbf{r}_{4}-\mathbf{r}_{i}|}}{| \mathbf{r}_{4}-\mathbf{r}_{i}|}. 
\end{equation}
The effective potential in Eq. \eqref{eqn::effective_potential} ensures that no artificial autoionization takes place. Indeed, when electron $i$ moves into the vicinity of the core, i.e., $ \mathbf{r}_{i}\rightarrow\mathbf{r}_{4},$ we have $V_{\mathrm{eff}}(\zeta_j,|\mathbf{r}_{4}-\mathbf{r}_{i}|) \to \zeta_j.$ Therefore, we have a finite energy transfer between the bound electrons $i$ and $j$.

The coefficients ${c_{i,j}(t)}$ control whether a given electron pair $(i,j)$ interacts through the full Coulomb potential or through the effective potentials $V_{\text{eff}}(\zeta_{i},|\mathbf{r}_{4}-\mathbf{r}_{j}|)$ and $V_{\text{eff}}(\zeta_{j},|\mathbf{r}_{4}-\mathbf{r}_{i}|)$ during time propagation. When at least one electron in the pair is quasifree, the coefficients ${c_{i,j}(t)}$ take their minimum value, i.e., 0, indicating that the full Coulomb potential is on. In contrast, when both electrons $i$ and $j$ are bound, the coefficients ${c_{i,j}(t)}$ take their maximum value, i.e., 1, indicating that the effective potentials $V_{\text{eff}}(\zeta_{i},|\mathbf{r}_{4}-\mathbf{r}_{j}|)$ and $V_{\text{eff}}(\zeta_{j},|\mathbf{r}_{4}-\mathbf{r}_{i}|)$ are on. To ensure that there is no discontinuity in the Hamiltonian when switching between the full Coulomb potential and the effective potentials, we set ${c_{i,j}(t)}$ to change linearly with time between the limiting values of zero and one \cite{Agapi3electron,PhysRevA.107.L041101},
\begin{equation}\label{eqn::zeta_charges_section}
{c_{i,j}(t)} = \left\{
    \begin{array}{ll}
        0 & {c(t)}  \leq 0 \\
        {c(t) } & 0 < {c(t)} < 1 \\
        1 & {c(t)}  \geq 1,
    \end{array}
\right.
\end{equation}
where ${c(t) = \beta (t-t^{i,j}_s)+c_{0},} $ and ${c_0}$ is the value of  ${c_{i,j}(t)} $ just before a switch at time ${t^{i,j}_s}$. The switch from full Coulomb to effective Coulomb potential occurs at time ${t^{i,j}_s}$ when at least one of the bound electrons in the pair $(i,j)$ changes to quasifree. Similarly, the switch from effective Coulomb potential to full Coulomb potential occurs at time ${t^{i,j}_s}$  if in a pair of a quasifree electron and a bound electron the quasifree electron becomes bound. At the start of the propagation, at time $t_0,$ we set ${t^{i,j}_s}=t_0$ and assign $c_0=1$ for electron pairs where both electrons are bound and $c_0=0$ otherwise. We have to ensure that the transition between interactions is slow enough so that there is no discontinuity in the Hamiltonian. At the same time this same transition must be fast enough so that when an electron becomes bound from quasifree there is no time for artificial  autoionization to take place.  These two conditions are satisfied by choosing $\beta=\pm 0.5$ which corresponds to a time interval of 2 a.u.;  the positive sign indicates a switch-on of the effective potential, while the negative sign indicates a switch-off.

The criteria we employ to distinguish between bound and quasifree electrons are thoroughly discussed and illustrated in Refs. \cite{Agapi3electron,43wt-x129}. For completeness, we provide a brief overview of the aforementioned criteria. A quasifree electron may become bound after a recollision takes place. Specifically, once the quasifree electron has passed its point of closest approach to the core, it is considered to be bound if the $z$ component of its position vector is influenced more by the core than the electric field \cite{Agapi3electron,43wt-x129}. Conversely, a bound electron can transition to quasifree due to transfer of energy during a recollision or due to the laser field. In the first case, a bound electron is labelled as quasifree if its potential energy with the core constantly decreases after recollision. In the second case, if the electron’s compensated energy becomes positive and stays positive the bound electron is registered as quasifree. We note that at high intensities an additional condition is needed. This extra condition is discussed in Section~\ref{Section:auto}.

We model the driving laser field using the vector potential
\begin{equation}\label{eq:vector_potential}
\mathbf{A}(y,t) = -\frac{E_0}{\omega}\exp \left[ - 2\ln (2)\left( \frac{c t - y}{c \tau} \right)^2 \right]   \sin ( \omega t  - k y) \hat{\mathbf{z}},
\end{equation}
with $k=\omega/c$ the  wave number of the laser field.  The parameters $E_0$ and $\omega$ denote the strength and radial frequency of the field, respectively, and $\tau$ is the full width at half maximum of the pulse duration in intensity. The direction of the vector potential and the electric field is along the $z$ axis, while the light propagates along the $y$ axis.  The corresponding magnetic field,
\begin{equation}
\mathbf{B}(y,t) = \nabla \times \mathbf{A}(y,t),
\end{equation}
is directed along the $x$ axis. 

\subsection{Initial conditions}
For the tunnel-ionizing electron (referred to as electron 1), we use  the dipole approximation to obtain the initial state at the start of the propagation. That is, we use the dipole approximation to obtain the tunneling rate and the exit point. The propagation start time, \( t_0 \), is randomly sampled using importance sampling~\cite{ROTA1986123} within the time interval \([ -2\tau, 2\tau ]\), during which the electric field is non-zero. The distribution we use for importance sampling  is  the nonrelativistic quantum mechanical tunneling rate described by the instantaneous Ammosov-Delone-Krainov (ADK) formula  \cite{Delone:91,Landau}, with empirical corrections  for high intensities, see Ref. \cite{Tong_2005}. The rate is also adjusted accordingly to account for the depletion of the initial ground state \cite{PhysRevA.107.L041101}. The core is initialized at rest at the origin. For the two initially bound electrons (electrons 2 and 3), we employ a microcanonical distribution~\cite{Agapi3electron}. 

In our previous studies \cite{Agapi3electron}, the exit point of the tunnel-ionizing electron at time $t_0$ was obtained analytically using parabolic coordinates \cite{Landau}, which are referred to as $\eta$ and $\xi$ in Section \ref{Section:e1_initial_con_a}. This exit point is computed under the assumption that $\eta(t_0) \gg \xi(t_0)$ \cite{Landau}. Already at 1.7 $\mathrm{PW/cm^2}$, we find that $\eta(t_0)$ is equal to roughly 4 times $\xi(t_0)$. Hence, we assume that $\eta(t_0) \gg \xi(t_0)$ is not satisfied for intensities equal to or higher than  1.7 $\mathrm{PW/cm^2}$. For these intensities, we specify the exit point numerically in Cartesian coordinates.  In what follows, we briefly describe  how to obtain the exit point of electron 1 with both methods.

\subsubsection{Exit point of electron 1 in parabolic coordinates}\label{Section:e1_initial_con_a}
We obtain the exit point analytically using parabolic coordinates, following the method outlined in Refs. \cite{Landau,Madsen2004}. For the tunnel-ionizing electron 1,  Schr\"{o}dinger's equation is given by 
\begin{align}\label{Hamiltonian}
H\psi= -I_{p,1}\psi \Rightarrow  -\dfrac{ \nabla^2\psi}{2} + V\psi + I_{p,1}\psi=0,
\end{align}
where
\begin{equation}
V = -\dfrac{Q_c}{r_{1}} + z_{1}E(t_0),
\end{equation}
with $\mathbf{r}_{1}=(x_{1},y_{1},z_{1})$ the position vector of electron 1, $Q_c$ the charge of the core, and $E(t_0)$ the electric field strength at time $t_0$. The energy of electron 1 is set equal to $I_{p,1}$, i.e., the first ionization potential. We employ parabolic coordinates, defined as follows \cite{Landau}
\begin{equation}
\xi={r_{1}}+{z_{1}}, \quad \eta={r_{1}}-{z_{1}},
\end{equation}
with
\begin{equation}
\quad {r_{1}}=\frac{1}{2}(\xi+\eta), \quad {z_{1}}=\frac{1}{2}(\xi-\eta).
\end{equation}

To solve \eq{Hamiltonian} in parabolic coordinates, one seeks eigenfunctions of the form \cite{Landau,Madsen2004}
\begin{equation}\label{Wavefunction}
\psi=\mathcal{N} \frac{{f}_1(\xi) {f}_2(\eta)}{\sqrt{\xi \eta}} {e}^{im\phi},
\end{equation}
with $m$ the magnetic quantum number and $\mathcal{N}$ a normalization constant. Substituting \eq{Wavefunction} in \eq{Hamiltonian}, expressed in parabolic coordinates, results in
\begin{align}
-\frac{1}{2} \frac{d^2 f_1(\xi)}{d \xi^2}+(U_1(\xi)-\varepsilon) f_1(\xi)&=0,\\
-\frac{1}{2} \frac{d^2 f_2(\eta)}{d \eta^2}+(U_2(\eta)-\varepsilon) f_2(\eta)&=0,
\end{align}
where the two potentials are given by
\begin{align}
U_1(\xi)&=-\frac{\beta_1}{2 \xi}+\frac{m^2-1}{8 \xi^2}+\frac{E(t_0) \xi}{8},\label{eq:ParabolicPotential_xi} \\ 
U_2(\eta)&=-\frac{\beta_2}{2 \eta}+\frac{m^2-1}{8 \eta^2}-\frac{E(t_0) \eta}{8}, \label{eq:ParabolicPotential_eta}
\end{align}
and the energy $\varepsilon$ of electron 1 is  $- I_{p,1} / 4$. The parameters $\beta_1$ and $\beta_2$ are given by 
\begin{align}
\beta_1 &= -\left[ \frac{\xi}{{f}_1(\xi)} \frac{{d}^2 {f}_1(\xi)}{{~d} \xi^2}+\left(-\frac{\xi {I}_{p,1}}{2}-\frac{{m}^2-1}{4 \xi}-\frac{  {E(t_0)} \xi^2 }{4}\right) \right]\\
\beta_2 &= - \left[ \frac{\eta}{{f}_2(\eta)} \frac{{d}^2 {f}_2(\eta)}{{~d} \eta^2}+\left(-\frac{\eta {I}_{p,1}}{2}-\frac{{m}^2-1}{4 \eta}+\frac{ {E(t_0)} \eta^2 }{4} \right) \right],
\end{align} 
and are connected via the constraint
\begin{equation}
\beta_1+\beta_2=Q_c.
\end{equation}

From Eqs. \eqref{eq:ParabolicPotential_xi} and \eqref{eq:ParabolicPotential_eta}, it follows that only $U_2(\eta)$ allows for tunneling. Hence, we focus on the solution of $U_2(\eta) = \varepsilon$ in order to specify the exit point.  This leads to the cubic equation
\begin{equation}\label{eq:qub2}
\eta^3  +a_2\eta^2 + a_1\eta + a_0  = 0,
\end{equation}
with
\begin{equation}
a_0 = -\frac{m^2-1}{E(t_0)}, \quad a_1 = \frac{4 \beta_2}{E(t_0)}, \quad a_2 = - \frac{2 I_{p,1} }{E(t_0)}.
\end{equation}
The solutions are given by \cite{abramowitz1964handbook}
\begin{align}\label{eq:sol_to_cubic_general}
\eta=\left\{\begin{array}{c}
\left({s}_1+{s}_2\right)-\frac{{a}_2}{3} \\
-\frac{1}{2}\left({s}_1+{s}_2\right)-\frac{{a}_2}{3}+{i} \frac{\sqrt{3}}{2}\left({s}_1-{s}_2\right) \\
-\frac{1}{2}\left({s}_1+{s}_2\right)-\frac{{a}_2}{3}-{i} \frac{\sqrt{3}}{2}\left({s}_1-{s}_2\right),
\end{array}\right.
\end{align}
where
\begin{equation}
s_1 = \left(r + \sqrt{q^3 + r^2} \right)^{\frac{1}{3}},
\end{equation}
and
\begin{equation}
s_2 = \left(r - \sqrt{q^3 + r^2} \right)^{\frac{1}{3}},
\end{equation}
with 
\begin{align}
q &= \frac{a_1}{3} - \frac{a^2_2}{9} \label{eq:q},\\
r &= \frac{a_1a_2 -3a_0}{6} -\frac{a^3_2}{27} \label{eq:r}.
\end{align}

For the below-the-barrier intensity regime, we find that $\mathrm{q^3 + r^2 < 0},$ and the three solutions of the cubic equations are real and take the form \cite{abramowitz1964handbook}
\begin{equation} \label{BelowBarrierEta}
\eta=\left\{\begin{array}{c}
2A\cos(\varphi/3) - \frac{a_2}{3}\\
-A \cos(\varphi/3) - \frac{a_2}{3} - \sqrt{3}A\sin(\varphi/3)\\
-A \cos(\varphi/3) - \frac{a_2}{3} + \sqrt{3}A\sin(\varphi/3),
\end{array}\right.
\end{equation}
where
\begin{align}
A^3 &= \sqrt{r^2 +  | q^3 + r^2 | }, \\
\varphi &= \arctan(\sqrt{|q^3 + r^2|}/r).
\end{align}
We find that the exit point of electron 1, corresponding to the solution furthest away from the core, is given by $\eta = 2A\cos(\varphi/3) - a_2/3$.

To obtain the parameters $\beta_1,\beta_2$, we assume that the exit point $z_1$ is large. In parabolic coordinates, this translates to $\eta \gg \xi$ and $ z_1(t_0) \approx - \eta(t_0)/ 2$ \cite{Madsen2004}. Then, the asymptotic Coulomb wave function is given by \cite{Madsen2004}
\begin{align}\label{eq:as_wavefunction}
& \psi \approx  \frac{2^{-(Q_c / \kappa)+1}}{|m|!} \xi^{|m| / 2} e^{-\kappa \xi / 2} \eta^{Q_c /  \kappa-|m| / 2-1 } e^{-\kappa \eta / 2} \frac{e^{i m \phi}}{\sqrt{2 \pi}},
\end{align}
where $\mathrm{\kappa=\sqrt{2I_{p,1}.}}$ We also assume that $E(t_0)/\kappa^3 \ll 1$ \cite{Landau,Madsen2004}. This condition is satisfied for all intensities used in this work. In this limit one finds that \cite{Madsen2004}
\begin{equation}\label{eq:beta1}
\beta_1 = \frac{( |m| + 1 ) \kappa }{2},
\end{equation}
and therefore,
\begin{equation}\label{eq:beta2}
\beta_2 = Q_c - \beta_1 = Q_c - \frac{( |m| + 1 ) \kappa }{2}.
\end{equation}
Here, we consider $\beta_1 = \beta_2 = 1/2$, as in our previous studies \cite{PhysRevA.78.023411} and other studies \cite{sarkadi_comparative_2021}.

For the below-the-barrier-regime, we take the initial momentum of electron 1 to be equal to zero along the laser field. The transverse momentum  is given by a Gaussian distribution.  The latter represents a Gaussian-shaped filter with an intensity-dependent width arising from standard tunneling theory \cite{Delone_1998,PhysRevLett.112.213001}.

For the over-the-barrier-regime, we find that $\mathrm{q^3 + r^2 > 0.}$ In this case, we place electron 1 at a distance $r_{\text {max }}$ that corresponds to the top of the barrier of $U_2(\eta).$ In addition, we set the magnitude of the momentum of electron 1, $p_{1}$, equal to \cite{PhysRevA.90.053419}
\begin{align}\label{Eq:momE1}
\begin{split}
p_{1} &=\sqrt{-2\left[I_{p,1}+V\left(r_{\max }, t_0\right)\right]}. 
\end{split} 
\end{align}
The direction of $p_1$ is uniformly distributed in space, with the only restriction being that $\mathbf{p}_{1} \cdot \mathbf{E}\left(t_0\right) \leqslant 0$, that is, electron 1 escapes opposite to the direction of the electric field.

Here, for the ionization potential of Ne, $I_{p,1}$, we find that the over-the-barrier regime corresponds to intensities higher than roughly 2.6 $\mathrm{PW/cm^2}.$ Also, as discussed above, for intensities higher than  1.7 $\mathrm{PW/cm^2}$, we do not employ parabolic coordinates.  Hence, in this work, parabolic coordinates are only employed in the below-the-barrier intensity regime.  For intensities higher than 1.7 $\mathrm{PW/cm^2}$, we obtain the exit point  of electron 1 numerically in Cartesian coordinates, as described in the next Section.

\subsubsection{Approximate exit point of electron 1 obtained in Cartesian coordinates }\label{Section:e1_initial_con_b}
\hfill\break 
We compute the  point  electron 1 exits from the field-lowered Coulomb potential barrier
using  equation
\begin{align}\label{ActualPot}
\begin{split}
V(\mathbf{r}_{1}, t_0) &= -\frac{Q_4}{r_{1}} + \sum_{j=2}^3 \frac{1}{|\mathbf{r}_{1}-\mathbf{r}_{j}|} + \mathbf{r}_{1} \cdot \mathbf{E}(t_0) \\
&= -I_{p,1}.   
\end{split}
\end{align} 
For the below-the-barrier intensity regime, we use Eq.\eqref{ActualPot} to specify $r_{1, \|}$, i.e., the component of $\mathbf{r}_{1}$ along  the field. To do so, we set equal to zero the component of $\mathbf{r}_{1}$  perpendicular to the field. As for parabolic coordinates, we set the momentum of  electron 1 along the field equal to zero. The transverse momentum is given by a Gaussian distribution \cite{Delone_1998,PhysRevLett.112.213001}.
 For the over-the-barrier intensity regime,  we place electron 1 at the top of the barrier, defined from Eq.\eqref{ActualPot}, and obtain the momentum as for parabolic coordinates.  

\subsubsection{Selection of exit point of electron 1}
\hfill\break
To ensure that we avoid any unphysical conditions, electron 1 has to be placed at least as far from the nucleus as specified by the exit point obtained by \eq{ActualPot}. The reason is that the potential in \eq{ActualPot} is used in the propagation of electron 1. At the same time, parabolic coordinates provide an exact solution of the Schr\"{o}dinger equation, while \eq{ActualPot} provides only an approximate one. Hence, for intensities above 1.7 $\mathrm{PW/cm^2},$ we specify the exit point of electron 1 using the potential provided in \eq{ActualPot}, since we cannot employ parabolic coordinates.  For intensities below 1.7 $\mathrm{PW/cm^2},$ we find the exit point using both methods described in Sections \ref{Section:e1_initial_con_a} and \ref{Section:e1_initial_con_b}. The method that results in an exit point further away from the nucleus is the one used to specify the exit point of electron 1. 

 \subsection{Addressing autoionization at high intensities}\label{Section:auto}
At high intensities, on average the exit point of electron 1 is closer to the nucleus compared to lower intensities. As a result, sometimes, we find that, before electron 1 is driven away from the nucleus by the electric field, it moves backwards towards the nucleus. This seems to also be due to a rapid change of the position of the bound electrons in the initial state. As discussed in Refs. \cite{Agapi3electron,PhysRevA.107.L041101}, we initialize electron 1 as quasifree, with the full Coulomb potential being turned on between this electron and the bound ones. However, having the full Coulomb potential being turned on when electron 1 moves backwards towards the nucleus can lead to unphysical autoionization. To prevent this, we implement an additional criterion  for electron 1  to transition from quasifree to bound.

 In our previous studies \cite{Agapi3electron,PhysRevA.107.L041101}, electron 1 transitions from quasifree to bound only if a recollision takes place. Here, we allow for such a transition to occur, even without a recollision, soon after the start of the propagation.
Namely, we check  whether electron 1, immediately after the start of  the propagation, moves backwards towards the nucleus and transitions from quasifree to bound. To do so, first, (i) we check whether the magnitude of the potential of the quasifree electron 1 with the nucleus, $V_{1,c},$ is greater than $V_{\text{min}}$. The potential  $V_{\text{min}}$  corresponds to a distance of 15 a.u. from the nucleus and  is used in our previous studies \cite{Agapi3electron,PhysRevA.107.L041101} to check whether an electron transitions from quasifree to bound  or vice versa. Then, we check whether (ii) $V_{1,c}$ is continuously increasing, i.e., $\frac{dV_{1,c(t_{n+5})}}{dt} > \frac{dV_{1,c(t_{n})}}{dt}$ for five $t_n$s which are five time steps apart, see Refs  \cite{Agapi3electron,PhysRevA.107.L041101}. If this is satisfied, then we consider electron 1 transitioning from quasifree to bound. If this is not satisfied at a time $t_n$, we re-initiate checking for condition (ii) at $t_n$. We stop checking condition (ii) if electron 1 becomes bound or if $V_{1,c} < V_{\text{min}}$. 

 \subsection{Tunneling during propagation}
\label{Sec:Tunnel}
In our previous studies of triple and double ionization of strongly driven atoms  \cite{Agapi3electron,PhysRevA.107.L041101}, we did not account for an electron tunneling during time propagation in the ECBB model. The reason is that the probability for sequential double ionization is very small for the intensities previously considered, up to 1.6 $\mathrm{PW/cm^2}$. We reach this conclusion by using a simple model of rate equations to describe single and sequential double ionization, see supplemental material of Ref. \cite{PhysRevA.107.L041101}. However, for intensities higher than  1.6 $\mathrm{PW/cm^2}$,  we find using this simple model that the probability for sequential double ionization is not negligible. Indeed, in Table \ref{tab:Probabilities}, we show that the  probability for sequential double ionization at  3 $\mathrm{PW/cm^2}$ is 7.6 \% (out of all ionization events).  All our results are obtained with a pulse of 800 nm wavelength and 25 fs duration.

We note that the probabilities and probability distributions (see next sections) obtained with the ECBB model when no tunneling is included are statistically equivalent to the ones obtained when considering only the events where no tunneling takes place in the  ECBB model with tunneling included. 

\begin{table}[h]
\centering
\begin{tabular}{lclclclclcl}
\hline
\multicolumn{5}{c}{Double ionization probability (\%)}\\
\hline
Intensity ($\mathrm{PW/cm^2}$ )& 1.87 & 2 & 2.31  & 3 \\
\hline
ECBB model w tunnel 	 &1.2 & 1.5  & 2.4 &7.0 \\
ECBB model w/o tunnel &0.8 & 0.9  & 1.0 &1.2  \\
Simple model 	     &0.4 & 0.6  & 1.7 &7.6 \\
\hline
\end{tabular}
\caption{Double ionization probability at intensities 1.87, 2, 2.31 and 3 $\mathrm{PW/cm^2}$ of strongly driven Ne, using  the ECBB model with tunneling during time propagation turned on (ECBB model w tunnel) and turned off  (ECBB model w/o tunnel). We also obtain sequential double ionization with the simple model comprised of rate equations, described in  Ref. \cite{PhysRevA.107.L041101}. }
\label{tab:Probabilities}
\end{table}

To account for the probability of sequential double ionization increasing with intensity, we allow, in the ECBB model, for any bound electron to tunnel during time propagation of strongly driven atoms. We account for tunneling in atoms in the same manner as we do for strongly driven two- and three-electron molecules \cite{PhysRevA.85.011402,PhysRevA.90.053419,PhysRevA.109.033106}. That is, we allow for a bound electron to tunnel at the classical turning points along the axis of the electric field using the Wentzel-Kramers-Brillouin (WKB) approximation \cite{WKB}. The transmission probability is given by \cite{WKB}
\begin{equation}\label{EQ:Notrootz}
\mathrm{T \approx \exp \left(-2 \int_{r_a}^{r_b}\left[2\left(V_{\mathrm{tun}}\left(r, t_{\mathrm{tun}}\right)-\epsilon_i\right)\right]^{1 / 2} d r\right)},
\end{equation}
with $V_{\mathrm{tun}}\left(r, t_{\mathrm{tun}}\right)$ the potential of the bound electron $i$ in the presence of the nuclei and the laser field, including also the effective potential terms \cite{PhysRevA.109.033106}, and $r$ the distance of the electron along the field.
 The energy of the electron at the time of tunneling, $t_{\text{tun}}$, is $\epsilon_i$  and $r_a$ and $r_b$ are the classical turning points. This additional quantum mechanical aspect of the ECBB model allows for sequential ionization. It allows for another electron to tunnel-ionize solely due to the laser field during time propagation, in addition to  electron 1  tunnel-ionizing at the start of the propagation.

In Table \ref{tab:Probabilities}, using the ECBB model with tunneling, we find the double ionization probability to be roughly equal to the one obtained with the ECBB model where  no tunneling is included plus the sequential double ionization probability given by the simple model. We note that at  3 $\mathrm{PW/cm^2}$ the double ionization probability obtained by the simple model as well as by the ECBB model including tunneling is 7\% and 7.6\%, respectively. These probabilities are much larger than the DI probability of 1.2\% obtained by the ECBB model when no tunneling is included. This clearly suggests that tunneling must be included in the ECBB model to account for the increased sequential DI at higher intensities $\mathrm{PW/cm^2}$. However, we note that, see  Table \ref{tab:Probabilities}, at 3 $\mathrm{PW/cm^2}$, the ECBB model with tunneling gives a double ionization probability that is less than adding the probabilities obtained by the ECBB model without tunneling and the simple model, unlike for smaller intensities. This suggests  that  the ECBB model with tunneling underestimates the contribution of sequential double ionization for intensities around 3  $\mathrm{PW/cm^2}$. We address this point again in Section \ref{Sec:Results}.

\subsection{Focal volume averaging}
In order to compare our theoretical results with experimental ones we implement focal volume averaging (FVA) using the method described in Ref. \cite{wang_disentangling_2005}. The need for FVA arises from experimental results depending on the distribution of laser intensities inherent in the Gaussian profile of the laser beam.

The FVA-distribution $f_{\alpha}^{  \mathrm{FVA} }$ of a variable $x$ for a process $\alpha$ for a peak intensity $I_0$ is given by
\begin{align}\label{EQ:GaussBeamFVA}
\begin{split}
    &f_{\alpha}^{  \mathrm{FVA} }\left(I_0,x\right) = \\
  &\int_0^{I_0} \frac{ \left[\int_{-\infty}^{\infty} \Gamma_{\mathrm{ADK}}(t,I) d t\right]}{I} \;P_{\alpha}(I) \; f_{\alpha}(I,x)  \;d I,
\end{split}
\end{align}
where $P_{\alpha}(I)$ is the probability for the process $\alpha$ to occur for an intensity $I$. For instance, $\alpha$ can denote triple or double ionization. This probability is given by
\begin{equation}\label{EQ:ProbabilityFVA}
    \begin{aligned}
& P_{\mathrm{\alpha}}(I)=\frac{N_{\mathrm{\alpha}}}{N} \\
\end{aligned}
\end{equation}
where $N_{\alpha}$ is the number of events corresponding to the process $\alpha$ and $N$ is the total number of ionization events. $\Gamma_{\mathrm{ADK}}(t,I)$ is the instantaneous ADK rate \cite{Delone:91} with corrections for high intensities \cite{Tong_2005}. The integral of the ADK rate over the time interval where the laser field is on is equal to the probability of electron 1 tunnel-ionizing. This probability divided by $I$ is the normalizing factor for the different intensities considered when we focal volume average. Also, $f_{\alpha}(I,x)$ is the distribution of the variable $x$ at an intensity $I$, normalised to one. For example, $x$ can be the sum of the momenta of the ionizing electrons along the direction of the field, see Figs. \ref{fig:SOM_DI_FVA}, \ref{fig:SOM_TI_FVA}, \ref{fig:PATHWAY_DI_FVA} and \ref{fig:PATHWAY_TI_FVA}.

To perform the focal volume averaging for the distributions presented in Section \ref{Sec:Results}, we consider the intensities shown in Table \ref{tab:intensities}. We include only a limited number of intensities,  since the computations are highly demanding. Indeed, the highest probability for triple ionization  at 3 $\mathrm{PW/cm^2}$, from the intensities considered here, is less than $10^{-3}$, i.e., very small. The intensity of 0.76 $\mathrm{PW/cm^2}$ was  the lowest one considered, since  the triple ionization probability is very small and effectively out of reach computationally for small intensities. Also, the intensities  listed in Table \ref{tab:intensities}, were chosen so that the change in the normalizing factor $\left[\int_{-\infty}^{\infty} \Gamma_{\mathrm{ADK}}(t,I) d t\right]/I$ between any two successive intensities is the same.

\begin{table}[t]
\centering
\begin{tabular}{c}
\hline
Intensity (PW/cm$^2$) \\
\hline
0.76 \\
0.85 \\
0.95 \\
1.08 \\
1.53 \\
1.87 \\
2.00 \\
2.31 \\
3.00 \\
\hline
\end{tabular}
\caption{List of laser pulse intensities used for performing focal volume averaging in this work.}
\label{tab:intensities}
\end{table}

\section{Results}\label{Sec:Results}
Next, using the extended ECBB model,  we present triple and double ionization spectra of strongly driven Ne and compare  with experimental results \cite{Zrost_2006,PhysRevLett.93.253001,Rudenko_2008}. We use a laser pulse with wavelength of 800 nm and duration of $\tau=25$ fs. For the results presented, we state explicitly whether focal volume averaging is considered or not. Also, in the following figures, the momenta are expressed in units of $\sqrt{U_p},$ where $U_p= E_0^2/ \left( 4 \omega^2 \right)$ is the ponderomotive energy, i.e., the average energy an electron gains from the laser field.

\subsection{Ratio of double to triple ionization probability}
\begin{figure}[t]
	\centering
\includegraphics[width=\linewidth]{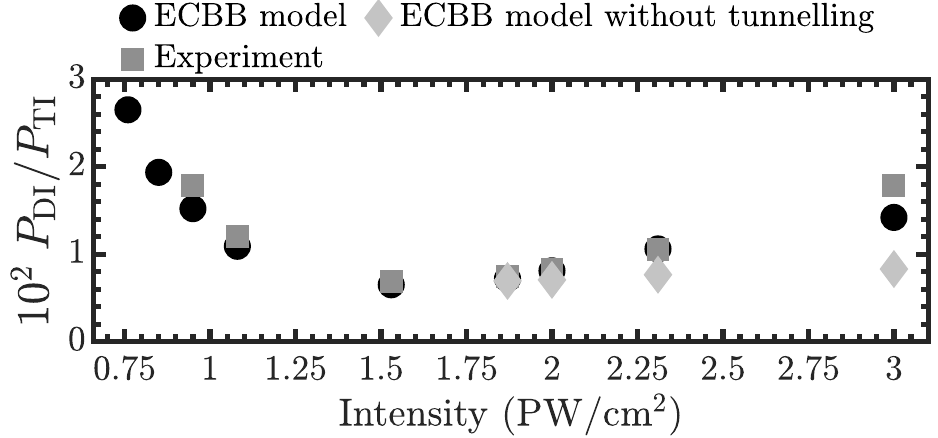}
\caption{For Ne, we plot the ratio of double to triple ionization probability obtained with the extended ECBB model (black circles), with the ECBB model when tunneling is not accounted for (light gray diamonds)  and compare with experiment (gray squares) \cite{Rudenko_2008}. These results are obtained without focal volume averaging.}
\label{fig:Probabilities}
\end{figure}
In \fig{fig:Probabilities}, for driven Ne, we compute the ratio of double to triple ionization probability using the extended ECBB model for all the intensities listed in Table \ref{tab:intensities}. Next, we compare with experiment \cite{Rudenko_2008} for the intensities where double and triple ionization yields are measured, i.e., in the interval ranging from 0.95 $\mathrm{PW/cm^2}$ to 3 $\mathrm{PW/cm^2}$, and find overall a very good agreement. At  3 $\mathrm{PW/cm^2},$ we find a difference of roughly 20 \% between our results and the experimental ones. This is most probably due to the extended ECBB model underestimating the probability for sequential double ionization, as discussed in Section \ref{Sec:Tunnel}, and overestimating sequential triple ionization, see next subsection.

\subsection{Distribution of the sum of the electron momenta} \label{Section::SOM}

\begin{figure}[b]
	\centering
\includegraphics[width=\linewidth]{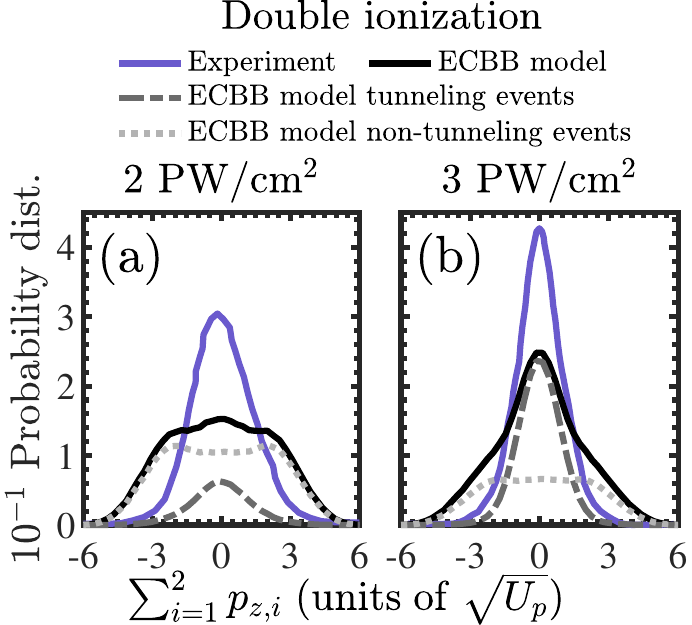}
\caption{For double ionization, we plot the probability distributions of the sum of the final electron momenta along the laser field  at intensities 2 PW/cm$^2$ (a) and 3 PW/cm$^2$ (b), accounting for focal volume averaging. Results are obtained with the ECBB model (black solid lines) and experimentally \cite{Zrost_2006,PhysRevLett.93.253001} (purple solid lines). Also, at each intensity, we separate the double ionization events depending on whether  an electron does or does not  tunnel  at least once   during time propagation. We plot  the respective distributions for double ionization events with tunneling    (light gray dotted lines) and without tunneling  (dark gray dashed-dotted lines). 
 The experimental and total theoretical distributions are normalized to one, while the distributions with and without tunneling events are normalized to their respective contribution to double ionization. }
\label{fig:SOM_DI_FVA}
\end{figure}

In Fig. \ref{fig:SOM_DI_FVA}, for double ionization, we plot the probability distribution of the sum of the final electron momenta along the laser  field, $p_z$, at intensities of 2 $\mathrm{PW/cm^2}$ and 3 $\mathrm{PW/cm^2}$.  We obtain our results using the extended ECBB model and account for  focal volume averaging. We find that the distribution  is much broader at 2 $\mathrm{PW/cm^2}$ compared to 3 $\mathrm{PW/cm^2}$. Also, the distribution at the higher intensity clearly peaks  around zero momentum.   We find the distributions obtained with the ECBB model (black lines) to be wider than the experimental ones (purple lines) \cite{Zrost_2006,PhysRevLett.93.253001}, which peak around zero momentum at both intensities.

   To better understand the shape of the momenta distributions in Fig. \ref{fig:SOM_DI_FVA}, at each intensity, we separate the double ionization events depending on whether or not an electron tunnels  at least once  during time propagation. For both subsets of double ionization events,  in the ECBB model, electron 1 tunnel-ionizes in the initial state, i.e., at the start of the propagation. We plot the respective distributions accounting for focal volume averaging. As expected, we find that the non-tunneling events (light gray dotted lines) give rise to the large width of the total momenta distributions (black lines). In contrast, the tunneling events (dark gray dashed-dotted lines), i.e., events that account for sequential double ionization, give rise to the peak around zero momentum of the  total momenta distributions (black lines).  At 3 $\mathrm{PW/cm^2}$, the percentage of tunneling and non-tunneling events is almost equal, consistent with the total distribution peaking mostly around zero momentum. Non-tunneling events dominate at 2 $\mathrm{PW/cm^2}$, consistent with the total distribution being significantly wider than at higher intensity. Given the above, the experimental distributions peaking more around zero momentum is consistent with the extended ECBB model underestimating sequential double ionization, particularly at 3 $\mathrm{PW/cm^2}$, see Section \ref{Sec:Tunnel}.

\begin{figure}[t]
\centering
\includegraphics[width=\linewidth]{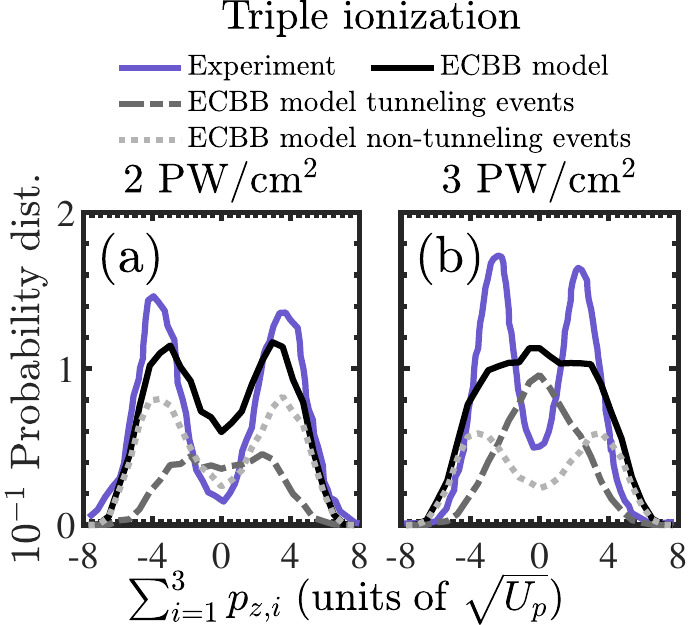}
\caption{For triple ionization, we plot probability distributions as in Fig.  \ref{fig:SOM_DI_FVA} and compare with experimental results   \cite{Rudenko_2008}.}
\label{fig:SOM_TI_FVA}
\end{figure}

Next, for triple ionization, in Fig. \ref{fig:SOM_TI_FVA}, we plot the probability distributions of the sum of the final electron momenta along the laser  field at 2 $\mathrm{PW/cm^2}$ and 3 $\mathrm{PW/cm^2}$. We find that the distribution at 2 $\mathrm{PW/cm^2}$ (black line) is wide and has a double peak  due to the non-tunneling events (light gray dotted line). This is expected, since non-tunneling events are  non-sequential multiple-ionization events and it is well known that NSMI events give rise to a double peak in momenta distributions \cite{PhysRevLett.87.043003,Becker_2005}. At 2 $\mathrm{PW/cm^2}$, our results agree very well  with experiment. At 3  $\mathrm{PW/cm^2}$, the double peak persists for the experimental distribution but less so for the one we obtain with the extended ECBB model. This difference at the high intensity might be due to the limited number of intensities we considered between 2 and 3 $\mathrm{PW/cm^2}$ in order to perform focal volume averaging. Indeed,Fig. \ref{fig:PATHWAY_PROB_TI}  shows that for all intensities considered below 3 $\mathrm{PW/cm^2}$ the pathways giving rise to a double hump structure prevail. 

The above analysis of the probability distributions of the sum of the final electron momenta reveals that including tunneling is important for accounting for sequential double ionization being significant at higher intensities. Indeed, 
for DI, the experimental distribution closely resembles the distribution obtained from the tunneling events in the ECBB model, see \fig{fig:SOM_DI_FVA}.
 In contrast, the experimental distribution for triple ionization resembles the distribution obtained  from the non tunneling events in the ECBB model, see \fig{fig:SOM_TI_FVA}. This suggests that the extended ECBB model overestimates sequential ionization for TI. Future studies could address this issue, for instance investigating how this behaviour is influenced by using different forms of effective potentials and their role in tunneling during propagation. 
  However, we note again that it is essential to include tunneling.  Indeed, the ratio of double to triple ionization  significantly deviates from the experimental values when tunneling is not included. In Fig. \ref{fig:Probabilities}, we show that, when tunneling is not taken into account,  
  at 2 $\mathrm{PW/cm^2}$ and 3 $\mathrm{PW/cm^2}$ the calculated ratio is approximately $25\%$ and $55\%$ smaller than the experimental value.

\subsection{Main pathways of three- and two-electron escape}

\begin{figure}[b]
\centering
\includegraphics[width=\linewidth]{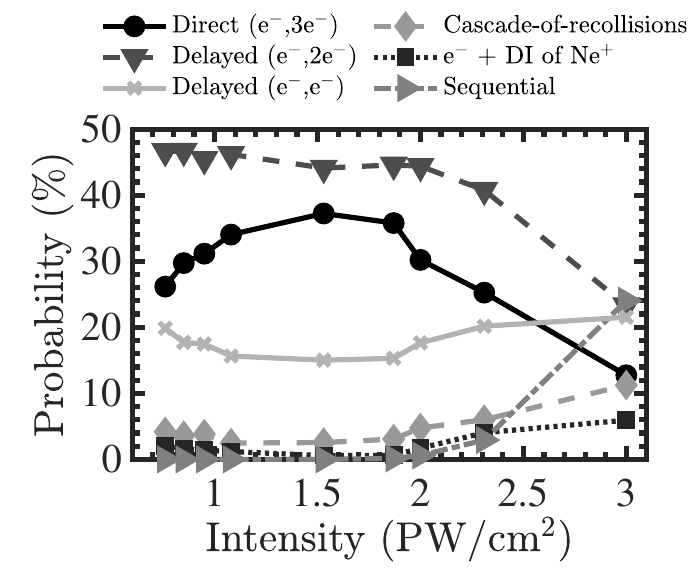}
\caption{Percentage contribution of the main pathways of triple ionization as a function of intensity.}
\label{fig:PATHWAY_PROB_TI}
\end{figure}

\begin{figure}[t]
\centering
\includegraphics[width=\linewidth]{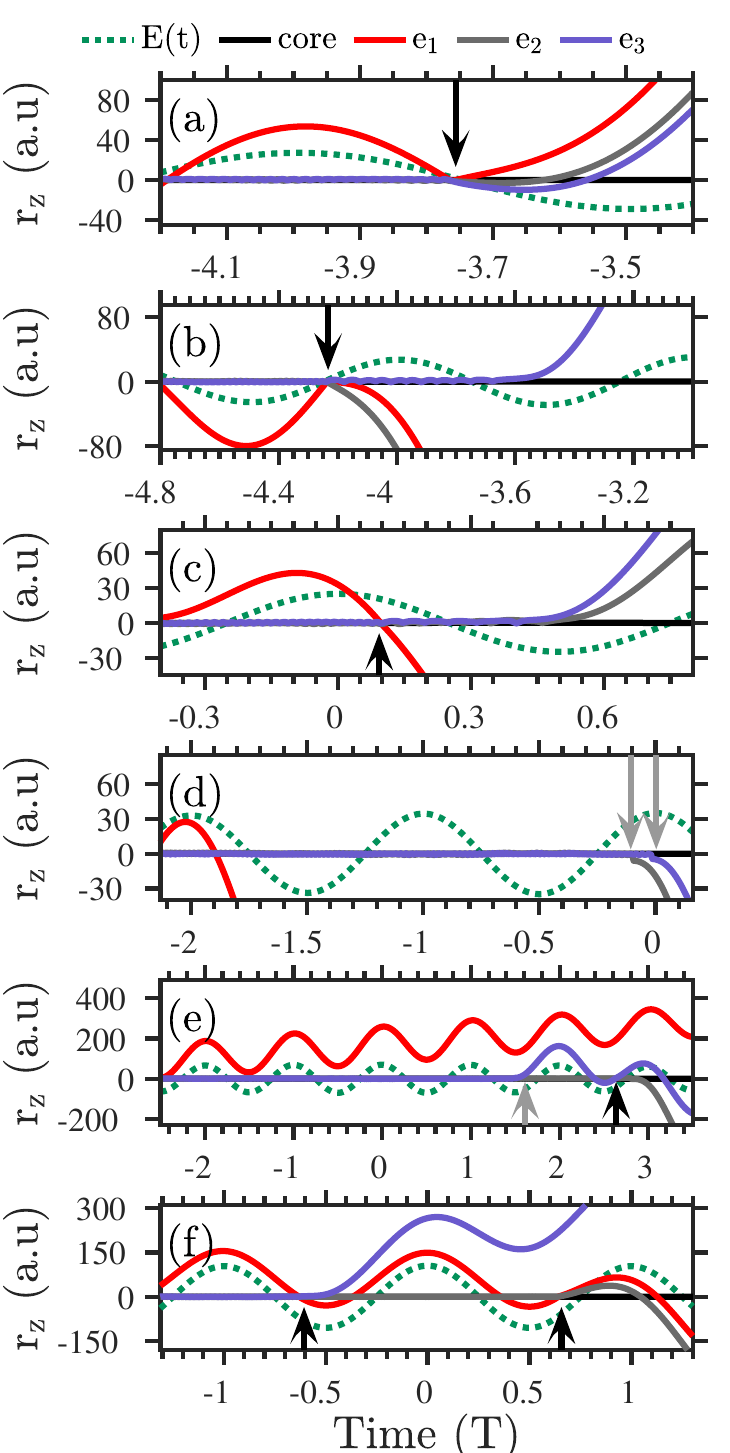}
\caption{Examples  of trajectories in  triple ionization pathways for Ne driven by a 25 fs, 800nm laser at 3 PW/cm$^2$. We plot the position of each of the three electrons along the field, electron 1 (red line), electron 2 (blue line) and electron 3 (gray line), while the electric field is depicted with a green dotted line.  Each  black arrow depicts the time of recollision between a quasifree electron and one or more bound electrons. A gray arrow indicates the time when an electron tunnels during time propagation. The pathways shown are direct (e$^-$, 3e$^-$) (a), delayed (e$^-$, 2e$^-$) (b),  delayed (e$^-$, e$^-$) (c), sequential (d), e$^{-}$ + DI of Ne$^{+}$ (e), cascade of recollisions (f).}
\label{fig:APPENDIX_PATHWAY}		
\end{figure}

Next, we identify the main pathways of  triple and double ionization in strongly driven Ne, see Ref. \cite{Agapi3electron} for a detailed description of how we identify pathways.  In Fig. \ref{fig:PATHWAY_PROB_TI}, we plot the percentage contribution  of the main pathways of triple ionization as a function of intensity. In all  pathways recollisions take place except for the sequential pathway where each electron escapes  due to the laser field.  During a recollision, an electron returns to the parent ion to transfer energy to the remaining electrons \cite{Corkum_1994}. For smaller intensities,  we find that the direct ($e^-$, 3$e^-$)  and the delayed ($e^-$, 2$e^-$) pathways prevail, as in our previous studies \cite{PhysRevA.107.L041101}. In the direct pathway,  all three electrons ionize soon after recollision, i.e., three highly correlated electrons escape, see Fig. \ref{fig:APPENDIX_PATHWAY}(a). Here, we refer to an electron escaping soon after recollision if the ionization time of this electron is within 1/8 T of the recollision time, with T the period of the laser field. In the delayed ($e^-$, 2$e^-$) pathway, the recolliding electron  transfers enough energy for  two electrons to ionize soon after recollision, resulting to two highly correlated electrons escaping. The other electron ionizes with a delay due to the field, see Fig. \ref{fig:APPENDIX_PATHWAY}(b).  The contribution of these two pathways reduces for intensities higher than  2 $\mathrm{PW/cm^2}$. 

For intensities higher than roughly   2.5 $\mathrm{PW/cm^2}$ the delayed ($e^-$, $e^-$) pathway becomes a prominent one along with the sequential one. In the delayed ($e^-$, $e^-$) pathway, the recolliding electron transfers enough energy to ionize only one electron soon after recollision, while  two electrons escape with a delay solely due to the field, see Fig. \ref{fig:APPENDIX_PATHWAY}(c). In the sequential pathway, no recollision takes place and the electrons ionize  sequentially solely due to the laser field, see Fig. \ref{fig:APPENDIX_PATHWAY}(d). In addition, at these higher intensities, we identify  two more pathways that contribute to triple ionization. Namely, in the 
cascade-of-recollisions pathway, the electrons escape due to the occurrence of two distinct recollisions, see Fig. \ref{fig:APPENDIX_PATHWAY}(f).  For the vast majority of these events,  the tunnel-ionizing electron 1 is the recolliding electron in both recollisions. For the higher intensity of  3 $\mathrm{PW/cm^2}$, we find that  for the vast majority of these events  each of the bound  electrons ionizes with a delay after each recollision. However, at 1.87 $\mathrm{PW/cm^2}$, in both recollisions,  each of the bound electrons has roughly the same probability  to ionize with a delay or soon after each recollision.
 Another pathway that contributes at higher intensities is the so-called  e$^{-}$ + DI of Ne$^{+}$
pathway. Electron 1 escapes  following  tunnel-ionization at the start of the propagation. In the remaining Ne$^+$  ion, one of the bound electrons tunnels and returns to the  ion to ionize the other bound electron, see Fig. \ref{fig:APPENDIX_PATHWAY}(e).  This latter electron ionizes mostly with a delay after recollision. 

\begin{figure}[t]
\centering
\includegraphics[width=\linewidth]{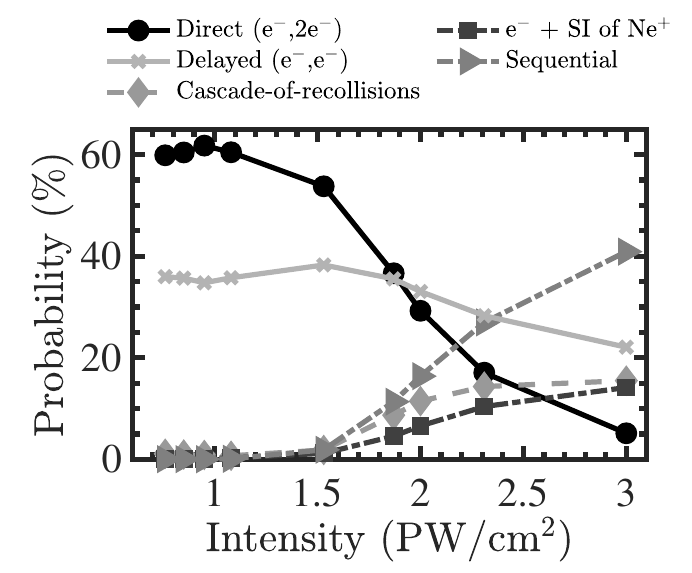}
\caption{Percentage contribution of the main pathways of double ionization as a function of intensity.}
\label{fig:PATHWAY_PROB_DI}
\end{figure}

In Fig. \ref{fig:PATHWAY_PROB_DI}, we plot the percentage contribution of the main pathways of double ionization as a function of intensity. We find that the direct ($e^-$, 2$e^-$) pathway prevails at small intensities, in agreement with our previous studies \cite{PhysRevA.107.L041101}, resulting in two highly correlated escaping electrons. Its contribution to double ionization decreases significantly with increasing intensity.  The delayed ($e^-$, $e^-$) pathway contributes less than the direct ($e^-$, 2$e^-$) pathway at small intensities. However, the contribution of the delayed ($e^-$, $e^-$) pathway is significant also at higher intensities. At higher intensities, roughly above 2.3 PW/cm$^2$, the sequential becomes the most prominent pathway. Moreover, two more pathways contribute at these higher intensities. One is the  cascade-of-recollisions pathway, where, as for triple ionization, two distinct recollisions take place. In the first recollision, electron 1 transfers energy and ionizes another bound electron, while in the second one electron 1 returns to the ion but does not ionize another electron. The other one is the    e$^{-}$ + SI of Ne$^{+}$, where electron 1 escapes  following its tunnel-ionization at the start of the time propagation; SI stands for single ionization. Also, one more electron tunnels due to the laser field. This latter electron returns to the ion to recollide but does not ionize the other bound electron.

\subsection{Contribution of the main pathways to the distribution of the sum of electron momenta} \label{Section::ResultsPathways}

\begin{figure}[t]
\centering
\includegraphics[width=\linewidth]{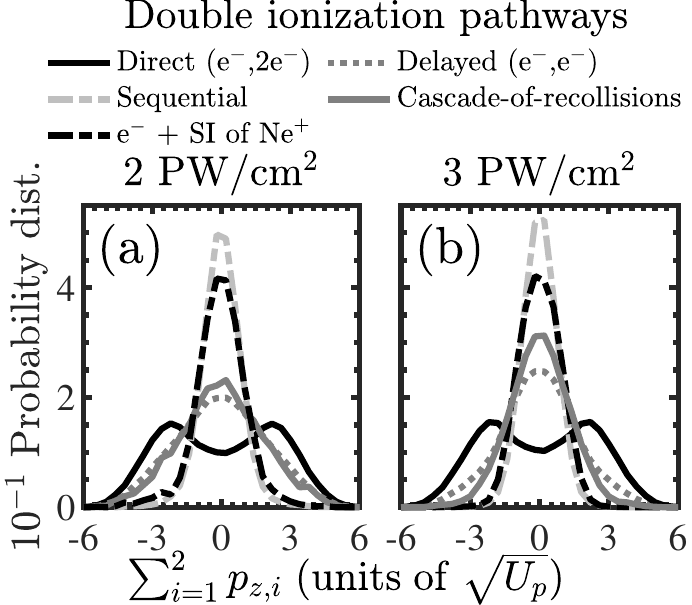}
\caption{For double ionization, probability distributions of the sum of the final electron momenta along the laser field at intensities 2 PW/cm$^2$, and 3 PW/cm$^2$ for the main pathways of two-electron escape. The focal volume average is included. Each distribution is normalized to one. }
\label{fig:PATHWAY_DI_FVA}
\end{figure}

Next, at 2 $\mathrm{PW/cm^2}$ and 3 $\mathrm{PW/cm^2}$, we identify  the contribution of the main pathways of double ionization (Fig. \ref{fig:PATHWAY_PROB_DI}) to the distribution of the sum of the final momenta $p_z$ of the two escaping electrons (Fig. \ref{fig:SOM_DI_FVA}). To do so, in 
Fig. \ref{fig:PATHWAY_DI_FVA}, we plot  the distributions of the sum of the final electron momenta along the field for each pathway of double ionization, while accounting for focal volume averaging. 
  In Fig. \ref{fig:PATHWAY_DI_FVA}, it is shown that the distribution of the direct  (e$^-$, 2e$^-$) pathway has a double peak structure. These peaks are located just short of $\pm$2  $\times \,$ 2$\sqrt{U_p}$. This is in accord with each of the two escaping electrons gaining momentum equal to $\mathrm{E_{0}/\omega=2\sqrt{U_p}}$ at the time of recollision, i.e., the maximum momentum they can gain from the laser field. For the delayed (e$^-$, e$^-$) and the cascade-of-recollisions pathways both distributions peak around zero momentum but are wide, while less so than the distribution of the   direct  (e$^-$, 2e$^-$) pathway. Indeed, in both pathways, the electron that escapes soon after recollision can gain up to   2$\sqrt{U_p}$ from the laser field, justifying why these distributions extend beyond this momentum. The  sequential and e$^{-}$+SI of Ne$^{+}$ pathways are the ones resulting in the least correlated electrons and hence their distributions peak sharply around zero momenta. The above features are present at both intensities, while at the higher one, with the exception of the direct pathway, all other distributions peak more sharply around zero momentum.
 Moreover, for the direct pathway we find that for the vast majority of events no tunneling is taking place during propagation. It is the other way around for the sequential and the e$^{-}$+SI of Ne$^{+}$ pathways. 
 
 Given the above, the distribution of the sum of the final electron momenta for all double ionization events at  2 PW/cm$^2$, shown in Fig. \ref{fig:SOM_DI_FVA}, is wide due to the direct pathway, mostly consisting of non-tunneling events. All other pathways, more so the sequential and e$^{-}$+SI of Ne$^{+}$ pathways consisting mostly of tunneling events, are responsible for the peak around zero. However, this peak is not pronounced     at  2 PW/cm$^2$, since for intensities up to 2 PW/cm$^2$ the direct pathway prevails.As the intensity increases, the direct pathway is no longer the prevailing one and the distribution of all other pathways peaks more sharply around zero momentum. This explains  why the distribution for all double ionization events at 3 PW/cm$^2$ peaks more sharply around zero compared to 2 PW/cm$^2$.

\begin{figure}[b]
\centering
\includegraphics[width=\linewidth]{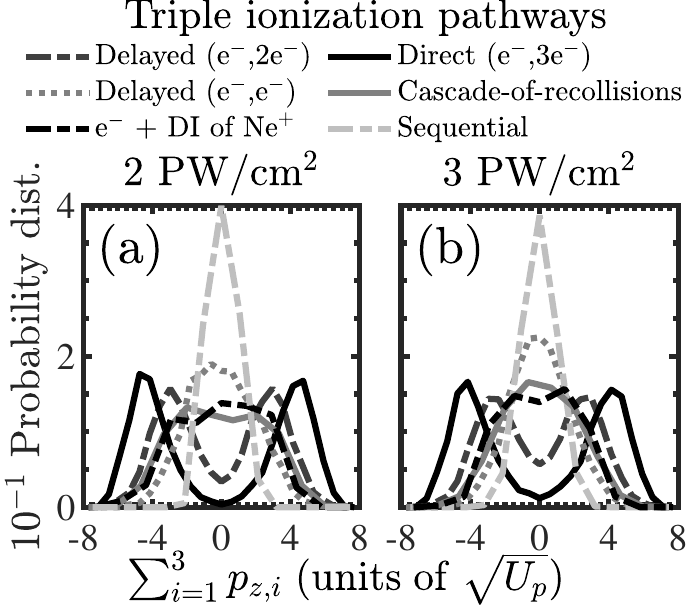}
\caption{For triple ionization, probability distributions of the sum of the final electron momenta along the laser field at intensities 2 PW/cm$^2$, and 3 PW/cm$^2$ for the main pathways of three-electron escape. The focal volume average is included. Each distribution is normalized to one. The distributions of  the e$^{-}$+DI of Ne$^{+}$ and cascades-of-recollisions pathways are not as smooth as for the other pathways due to the lower statistics for these pathways combined with these distributions  being wide. }
\label{fig:PATHWAY_TI_FVA}
\end{figure}

Next, at 2 $\mathrm{PW/cm^2}$ and 3 $\mathrm{PW/cm^2}$, we identify  the contribution of the main pathways of triple ionization (Fig. \ref{fig:PATHWAY_PROB_TI}) to the distribution of the sum of the final momenta $p_z$ of the two escaping electrons (Fig. \ref{fig:SOM_TI_FVA}). To do so, in 
Fig. \ref{fig:PATHWAY_TI_FVA}, we plot  the distributions of the sum of the final electron momenta along the field for each pathway of triple ionization, while accounting for focal volume averaging. 
  In Fig. \ref{fig:PATHWAY_TI_FVA}, it is shown that the distributions of the direct  (e$^-$, 3e$^-$) and delayed (e$^-$, 2e$^-$) pathways have a double peak structure. For the direct pathway, these peaks are located at $\pm$3  $\times \,$ 2$\sqrt{U_p}$. This is in accord with each of the three escaping electrons gaining momentum equal to 2$\sqrt{U_p}$, since they escape soon after recollision. For the delayed pathway,  these peaks are located at $\pm$2  $\times \,$ 2$\sqrt{U_p}$, since   two electrons can escape with momentum equal to 2$\sqrt{U_p}$ soon after recollision. The distributions of the e$^{-}$+DI of Ne$^{+}$ and cascades-of-recollisions pathways are still wide but less so than the distribution for the delayed (e$^-$, 2e$^-$) pathway.   The significant width of the distributions of these pathways   is mainly due to the significant percentage of events where  two electrons escape soon after recollision. This percentage increases to more than half at 1.87   $\mathrm{PW/cm^2}$, explaining why the distributions are more wide at 2 $\mathrm{PW/cm^2}$. In the delayed  (e$^-$, e$^-$) and, particularly, in the sequential pathway the escaping electrons are the least correlated giving rise to distributions peaking sharply around zero momentum. For the vast majority of events in the direct pathway there is no tunneling. It is the other way around for the sequential, delayed   (e$^-$, e$^-$),  e$^{-}$+DI of Ne$^{+}$ and cascades-of-recollisions pathways.

 Given the above, the distribution of the sum of the final electron momenta for all triple ionization events at  2 PW/cm$^2$, shown in Fig. \ref{fig:SOM_TI_FVA}, is wide mainly due to the direct (e$^-$, 3e$^-$) and delayed (e$^-$, 2e$^-$) pathways, mostly consisting of non-tunneling events. The sequential and delayed (e$^-$, e$^-$) pathways, consisting mostly of tunneling events, prevail at higher intensities, in accord with the distribution for all triple events having no double peak structure at 3  PW/cm$^2$, shown in Fig. \ref{fig:SOM_TI_FVA}. However, the distribution for all triple ionization events is still wide at 3  PW/cm$^2$ due to the large width of the distributions of the direct (e$^-$, 3e$^-$) and delayed (e$^-$, 2e$^-$) pathways, and due to the distributions of the e$^{-}$+DI of Ne$^{+}$ and cascades-of-recollisions pathways being less but wide nevertheless.
 
\section{Conclusions}
In conclusion, to address triple and double ionization of strongly driven Ne at higher intensities of 2 and 3 $\mathrm{PW/cm^2}$,  we extend the 3D ECBB semiclassical model. This extension concerns changing the initial conditions of the tunnel-ionizing electron, adding one more criterion to define a transition from quasifree to bound, following immediately after the start of the time propagation, and allowing electrons to tunnel during time propagation. Using this extended ECBB model, we obtain the distribution of the sum of the final electron momenta along the field for triple and double ionization. Comparing with experiments, we find a better agreement for the triple rather than the double ionization spectra. We attribute this to the ECBB model underestimating the probability for sequential double ionization at higher intensities. At 3 $\mathrm{PW/cm^2}$, we find that the ECBB model overestimates sequential ionization in triple ionization. To improve the extended ECBB model, future studies should address the role of the effective potential in tunneling and hence in sequential ionization. Also, we identify the main pathways of triple and double ionization as a function of intensity and identify pathways that significantly contribute only at higher intensities. Finally, we explain how each of these pathways gives rises to the main features of the distributions of the sum of the final electron momenta. We believe that this current work  further demonstrates that the ECBB model offers promising avenues for enhancing our theoretical understanding of non-sequential multiple ionization. The current work also constitutes a step into bridging the gap between theoretical studies  and experiments.

\section*{Acknowledgments}
S.J.P., G.P.K. and A.E. acknowledge EPSRC Grant No. EP/W005352/1. The authors acknowledge the use of the UCL Myriad High Performance Computing Facility (Myriad@UCL), the use of the UCL Kathleen High Performance Computing Facility (Kathleen@UCL), and associated support services in the completion of this work.

\section*{References}
\bibliographystyle{iopart-num}
\bibliography{bibliography}

\end{document}